\documentclass[12pt,preprint]{aastex}

\begin{document}

\title{FRACTAL DIMENSION OF INTERSTELLAR CLOUDS:
OPACITY AND NOISE EFFECTS}

\author{N\'estor S\'anchez,\altaffilmark{1,2}
        Emilio J. Alfaro,\altaffilmark{1} and 
        Enrique P\'erez\altaffilmark{1}}
\altaffiltext{1}{Instituto de Astrof\'{\i}sica de Andaluc\'{\i}a,
                 CSIC, Apdo. 3004, E-18080, Granada, Spain;
                 nestor@iaa.es, emilio@iaa.es, eperez@iaa.es.}
\altaffiltext{2}{Departamento de F\'{\i}sica, Universidad del
                 Zulia, Maracaibo, Venezuela.}

\defcitealias{san05}{Paper~I}
\defcitealias{san06}{Paper~II}

\slugcomment{The Astrophysical Journal: accepted}

\begin{abstract}
There exists observational evidence that the interstellar medium
has a fractal structure in a wide range of spatial scales. The
measurement of the fractal dimension ($D_f$) of interstellar clouds
is a simple way to characterize this fractal structure, but  several
factors, both intrinsic to the clouds and to the observations,
may contribute to affect the values obtained. In this work
we study the effects that opacity and noise have on the determination
of $D_f$. We focus on two different fractal dimension estimators: the
perimeter-area based dimension ($D_{per}$) and the mass-size  dimension
($D_m$). We first use simulated fractal clouds to show that opacity
does not affect the estimation of $D_{per}$. However, $D_m$ tends to
increase as opacity increases and this estimator fails when applied
to optically thick regions. In addition, very noisy maps can seriously affect
the estimation of both $D_{per}$ and $D_m$, decreasing the final estimation
of $D_f$. We apply these methods to emission maps of Ophiuchus, Perseus
and Orion molecular clouds in different molecular lines and we obtain
that the fractal dimension is always in the range 
$2.6 \lesssim D_f \lesssim 2.8$ for these regions.
These results support the idea of a relatively  high
($> 2.3$) average fractal dimension for the interstellar medium,
as traced by different chemical species.
\end{abstract}

\keywords{ISM: clouds --- ISM: individual (Ophiuchus, Orion,
          Perseus molecular cloud) --- ISM: structure}

\section{INTRODUCTION}

For a complete understanding of the physical processes involved
in the structure and evolution of the interstellar medium (ISM)
it is essential to characterize systematically this structure.
A systematic and uniform analysis would probably allow to draw
reliable conclusions on the ``real" ISM structure as well as
its dependence on variables such as galactocentric
distance or star formation activity. A simple approach consists
of characterizing the ISM topology through its fractal dimension.
Observations show that the boundaries of interstellar clouds have
projected dimensions ($D_{per}$) that are always in the range
$1.2 \lesssim D_{per} \lesssim 1.5$. This seems to be valid for IRAS
cirrus \citep{baz88}, molecular clouds \citep{dic90,fal91,lee04},
high-velocity clouds \citep{vog94}, H I distribution \citep{wes99},
etc. The general belief is that $D_{per}$ has a more or less
universal value around $\sim 1.35$, and this result could have
important implications because it is reasonable to assume that
clouds subject to the same underlying physical processes should
have the same fractal dimension. However, often the observational
data and/or analysis techniques are so different that the
robustness of this conclusion is questionable.

In a previous work \citep[hereafter \citetalias{san05}]{san05}
we showed that if the boundary of a projected cloud had
dimension $D_{per} \simeq 1.35$ then the three-dimensional
fractal dimension would be $D_f \simeq 2.6$, a value higher
than the value $D_f = D_{per} + 1 \simeq 2.35$ sometimes
assumed in the literature \citep[e.g.][]{elm96}.
Moreover, the average properties of the ISM are in
gross agreement with relatively high $D_f$ values
\citep[hereafter \citetalias{san06}]{san06}.
The application of two different fractal dimension estimators
(the perimeter and the mass dimensions) to Orion A molecular
cloud yielded $D_f \sim 2.6 \pm 0.1$ for this region. In
this work we apply the same techniques to various molecular
cloud maps, in a very first attempt of systematically comparing
fractal properties in different regions
and from different emission lines
of the ISM. An important point to take into account
is the sensitivity of these measurements to factors
such as finite sampling of the maps, resolution, noise, etc.
In \citetalias{san05} we showed that low resolution maps tend 
to decrease the estimated value of $D_{per}$.
The analysis of clouds mapped in different emission
lines opens up the question of the role played by self-absorption
in the estimation of the fractal dimension of the clouds. It
is obvious that what we observe is not only a projected image
of the true three-dimensional cloud but also a fraction of the
total emission of the cloud. Particles closer to the observer
will hide  --for some particular combinations of size, geometry
and absorption coefficients--  the emission coming from the
back side of the clouds. How much self-absorption is affecting
the estimation of the fractal dimension of the cloud? In
Section~\ref{sec_opacity} we analyze the effect that different
opacities would have on the measured $D_{per}$ and $D_m$ values.
After that, in Section~\ref{sec_application}, we use different
emission maps to calculate the fractal dimension of three
different molecular clouds (Ophiuchus, Perseus and Orion).
As a natural consequence of this analysis the
signal-to-noise ratio arises as an important factor contributing
to the uncertainty in the final estimation. This issue is
discussed in Section~\ref{sec_noise} where three different
views (three different transitional lines) of the same cloud
are analyzed for evaluating the fractal dimensions.
Finally, the main conclusions are summarized in
Section~\ref{sec_conclusions}.

\section{OPACITY EFFECT ON THE ESTIMATION OF THE FRACTAL DIMENSION}
\label{sec_opacity}

\begin{figure}
\plotone{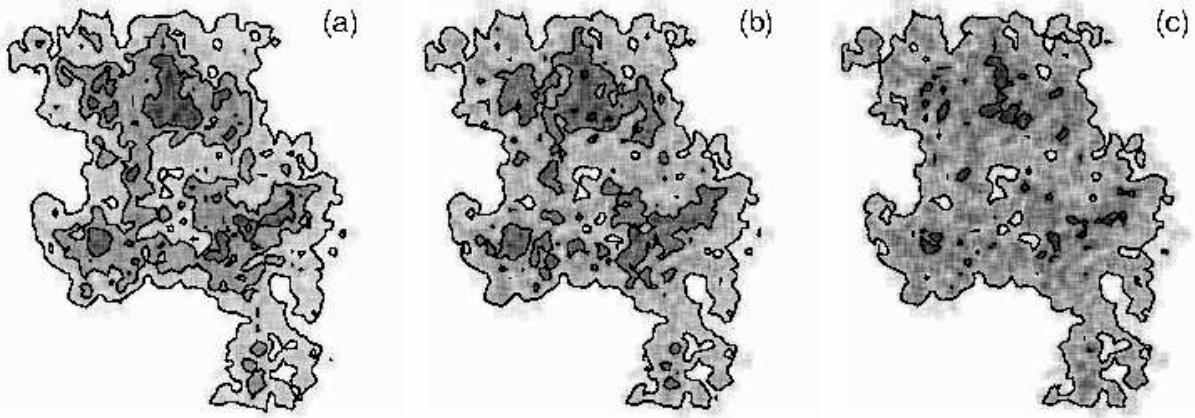}
\caption{Three images projected of the same cloud with
fractal dimension $D_f=2.6$ but for three different
optical depth values: (a) $\tau_0=0$, (b) $\tau_0=1$ and
(c) $\tau_0=2$. The contour levels are fixed at 25\%,
50\% and 75\% of the maximum projected intensity for
the case $\tau_0=0$.}
\label{example}
\end{figure}

We have generated fractal distributions of points by randomly
placing spheres inside spheres through a given number of levels
of hierarchy. In addition we have used a Gaussian kernel to
calculate the three-dimensional density field $\rho (x,y,z)$
associated to the fractal cloud. We refer readers to
\citetalias{san05} and \citetalias{san06} for details about
the procedure used.
In \citetalias{san05} we considered that the contribution
of every point to the projected image was the same, regardless
how they were distributed inside the cloud. In other words,
every particle acts as a similar emitter and what we observe
at every surface pixel is the summation of all the particles
projected on it, so the cloud is effectively optically thin.
Now we try to give a more realistic view of the projected 
cloud accounting for opacity effects in the
cloud. Thus, the observed emission of every particle is not
the same and depends on the column density that radiation
has to cross before exiting the cloud.
We have modified the algorithm in such a way that
contributions are weighted by $\exp\left[-\tau(x,y)\right]$
when the projection is done on, for example, the plane
$z=z_0$, being
\begin{equation}
\tau(x,y) = c \int_{z_0}^{z} \rho (x,y,z) dz
\end{equation}
the total optical depth between the point $(x,y,z)$ and the
projection plane. The absorption constant $c$ includes
quantities such as the abundance, mean molecular weight and
absorption cross-section of the emitting molecule, which we
assume constant throughout the structure. For the sake
of clarity, we will use the constant $\tau_0$: the maximum
optical depth in the case in which all the mass ($M_f$) is 
homogeneously distributed throughout the entire available 
volume ($V_f = (4/3) \pi R_f^3$). Since we have defined
$M_f=1$ and $R_f=1$, we obtain $\tau_0 = 3c/(2\pi)$. As
an example, Figure~\ref{example}
shows three projected images of the same cloud with fractal
dimension $D_f=2.6$ but for three different maximum optical
depth values ($\tau_0=0$, $1$ and $2$). The total optical
depth is in general a function of the position
in the projected map, but its maximum
value is always close to $\tau_0$. For the example shown
the maximum optical depth is $\sim 0.9$ and $\sim 1.7$ when
$\tau_0=1.0$ and $2.0$, respectively.

\begin{figure}
\plotone{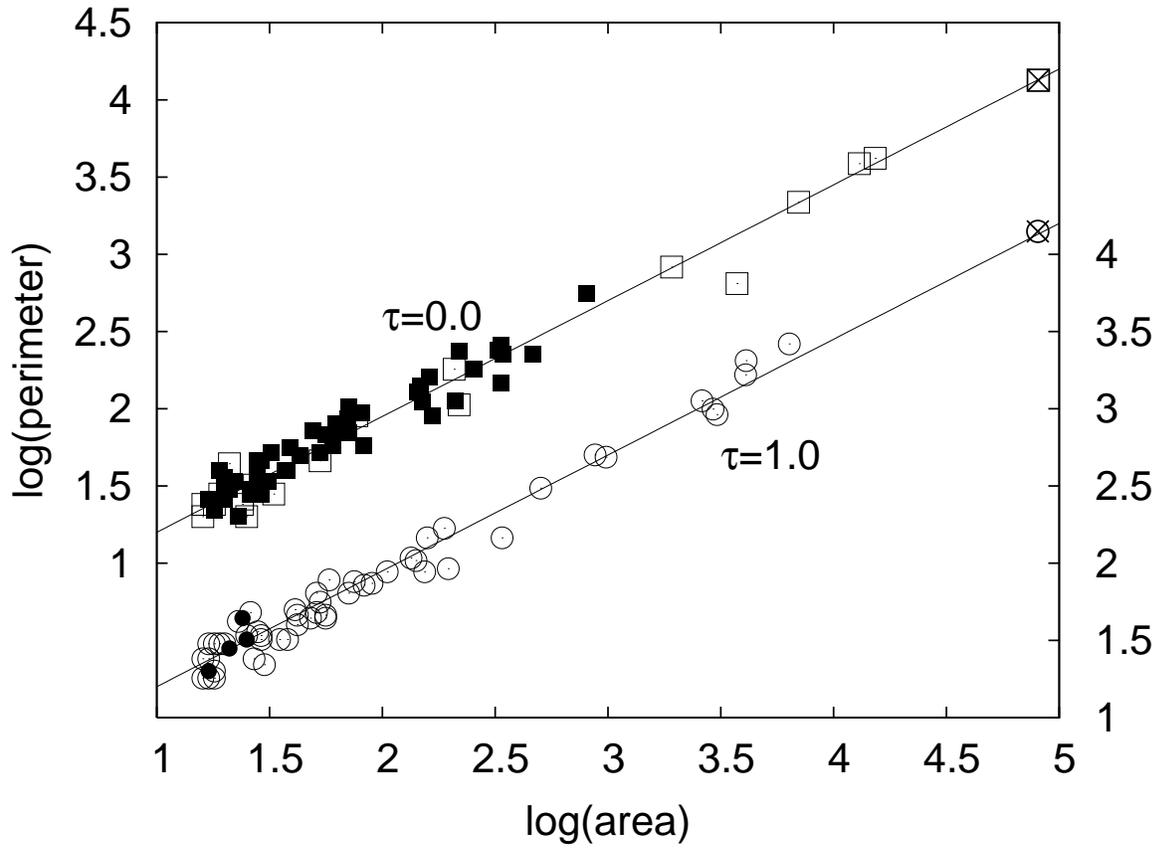}
\caption{ The perimeter as a function of the area for the
same fractal cloud shown in Figure~\ref{example}. Squares
(left axis) are for the case $\tau_0=0$ and circles (right
axis) for $\tau_0=1$. The intensity levels are fixed at
25\% (crossed symbols), 50\% (open symbols) and 75\% (filled
symbols) of the maximum intensity on the image.}
\label{periarea}
\end{figure}

As we can note in Figure~\ref{example}, the main effect of
opacity is to shorten the dynamical range of intensity levels
as well as to decrease the emission maxima.
Here we want to understand how this effect alters the
estimation of $D_f$. To do this we will use the two estimators
used in \citetalias{san05}: the perimeter-area based dimension
($D_{per}$) and the mass dimension ($D_m$). The
first method begins fixing a threshold intensity level
and defining each object as the set of connected pixels
whose intensity value is above this threshold.
Then the perimeter and the area of each object in the
image is determined and the best linear fit in a
log(perimeter)-log(area) plot is calculated. The slope
of this fit is $D_{per}/2$ \citep{man83}. To increase
the number of data points in the linear fit it is useful
to take several intensity levels. The second method
($D_m$) works by generating random positions along the
image and then placing cells of different radii (see
details in \citetalias{san05}). The ``mass" of each
cell is assumed to be the summed values of all the
intensities, and $D_m$ is calculated as the slope
of the best linear fit in a log(mass)-log(radius) plot.
We have run exactly the same algorithms as in \citetalias{san05}
to calculate $D_{per}$ and $D_m$ for several random fractal
clouds and random projections with different opacities. Our
first result is that the mean value of $D_{per}$ is not
significantly affected by the cloud opacity. The results
for $D_f=2.0$, $2.3$ and $2.6$ are shown in Table~\ref{table1},
where we can see that $D_{per}$ stays always within the standard
deviation independently of the opacity. For
a better understanding of this important result
Figure~\ref{periarea} shows, as an example, the
log(perimeter)-log(area) plot resulting from using only
three intensity levels (0.25, 0.5 and 0.75 times the
maximum projected intensity $I_{max}$) for the same
fractal cloud shown in Figure~\ref{example}. Squares
in Figure~\ref{periarea} refer to the case $\tau_0=0$
and circles to the case $\tau_0=1$. For the lowest
intensity level ($0.25 I_{max}$, denoted as crossed
symbols) only one relatively large structure (area
$\sim 10^5\ pixel^2$) is observed, which represents
the whole molecular complex. As the threshold intensity
is increased, smaller and denser structures
which are ``embedded" in the complex can be observed
(see Figure~\ref{example}). The central and densest
parts (cores) corresponding to a threshold intensity
of $0.75 I_{max}$ (filled symbols) are difficult to
detect for the case $\tau_0=1.0$, because opacity
occults the internal structure of the densest regions.
In contrast, small and low-density regions as well as
the gas that lies near the boundaries of the
three-dimensional clouds are less affected because
they have relatively low column densities. However,
the same linear behavior is found for different $\tau_0$
values (slopes in Figure~\ref{periarea} are similar
within the fit errors) because the ideal
monofractal clouds we are simulating keep the same
fractal properties at all the spatial scales considered,
i.e., the fractal dimension is the same for both large
low-density clouds and small high-density cores. This
is the reason why the perimeter-area dimension remains
almost unchanged. Thus, $D_{per}$ appears as a robust
estimator of the fractal dimension given that the shape
of the external contour is not modified by opacity, rather
it is mainly determined by the internal structure of the
cloud. The measure of $D_{per}$ can, in this way, be used
to infer the fractal structure of the cloud regardless the
opacity of the observed transition line.

The situation is different for the mass dimension because
this estimator has to use, unlike $D_{per}$, information
from all the cloud structure (mass versus radius) to
quantify $D_m$, including the internal and dense regions
which could be hidden in the projected image due to opacity
effects. The results for mass dimension are also shown in
Table~\ref{table1}. For the particular case $D_f=2.6$ we
observe small but significant variations with opacity. The
general trend is to increase $D_m$ as $\tau_0$ increases,
an expected result taking into account the fact that higher
$\tau_0$ values produce maps with shorter dynamical range
of intensities. But additionally the errors
become higher (worse mass-size correlation) and the
method begins to fail (correlation not found) for
$\tau_0 \gtrsim 1.3$.

\section{APPLICATION TO MOLECULAR CLOUD MAPS}
\label{sec_application}

Considering that opacity almost does not affect
the estimation of the perimeter-area dimension for the
simulated fractal clouds, we set out to study the
fractal dimension of nearby interstellar clouds mapped
in different molecular lines. As starting hypothesis,
we argue that if different molecules are distributed
following very similar patterns then their maps should
exhibit nearly the same perimeter-area dimension values,
independently of the opacity of the molecular transition
line. On the opposite, statistically significant differences
will be evidence of internal structure differences.
We have used various maps of molecular clouds to calculate
both $D_{per}$ and $D_m$. We have searched
the literature for available similar
maps observed in different molecular lines. We first use
integrated intensity maps of Ophiuchus and Perseus molecular
clouds obtained from the COMPLETE Survey of Star-Forming
Regions \citep{rid06}. The maps were obtained from simultaneous
observations in the $^{12}$CO 1-0 and $^{13}$CO 1-0 transitions
at the 14m Five College Radio Astronomy Observatory (FCRAO).
The half-power beamwidth (HPBW) is around 45\arcsec\ for both lines,
the data are over-sampled at irregular intervals and they were
convolved onto a regular 23\arcsec\ grid. We have also used integrated
intensity maps of Orion molecular cloud obtained from observations
with the 45m telescope of the Nobeyama Radio Observatory
\citep{tat93}. We use three maps of the region around Orion
KL in the $^{13}$CO 1-0, CS 1-0 (observed simultaneously) and
C$^{18}$O 1-0 transitions. The HPBW was 36\arcsec\ (for CS)
and 15\arcsec\ (for $^{13}$CO and C$^{18}$O)
with a grid spacing of 40\arcsec\ (CS and $^{13}$CO) and
$\sim$34\arcsec\ (C$^{18}$O). After re-gridding
the maps have resolutions of 10\arcsec\ (CS and $^{13}$CO)
and 17\arcsec\ (C$^{18}$O). In principle, each map provides
important information on cloud structure. The high $^{12}$CO
abundance ensures strong emission occurring throughout most of
the structure, but the lower-J lines of this molecule are often
optically thick providing very little information on the
structure of very dense regions within molecular clouds. On
the opposite, the lines of lower abundance molecules (such
as, for instance, C$^{18}$O) are usually optically thin even
on multi-parsec scales making them suitable for identifying
deep regions, but the emission is limited to the denser gas.

The results are summarized in Figures~\ref{perimeter} and
\ref{mass} which show the perimeter and mass dimensions, respectively,
obtained for each of the maps (the bars on the data points
are one standard deviation resulting from the best linear
fit in the perimeter-area or mass-size log-log plot, see
\citetalias{san05}). The perimeter-area method always gives
three-dimensional fractal dimensions in the range
$2.6 \lesssim D_f \lesssim 2.8$ for the Ophiuchus, Perseus
and Orion molecular clouds. The exception to this general
result was the C$^{18}$O map of Orion, which will be discussed
in the next section. For each molecular cloud the $D_{per}$ value
does not depend, within the error bars, on the transition line used,
a behavior which is consistent with the results we found in
Section~\ref{sec_opacity}.
The mass-size method yields $2.5 \lesssim D_f \lesssim 2.8$
for all the maps (except again the C$^{18}$O map),
which is in gross agreement with the perimeter-area
method. However, here we obtain higher error bars doing
more difficult to constraint the range of $D_f$ values.
Part of this uncertainty is associated with the method
itself but part is due to its sensitivity to opacity
(Section~\ref{sec_opacity}), because opacity variations
within each map will affect the looked for correlation.
In spite of this limitation, the mass-size method is
usefull as an additional and independent tool for
verifying values and trends derived from the
perimeter-area method, specially in low opacity
regions. An example is the relatively low fractal
dimension value for the original C$^{18}$O map
which is obtained from both $D_{per}$
and $D_m$, and it will be discussed next.

\section{THE EFFECT OF NOISE}
\label{sec_noise}

\begin{figure}
\plotone{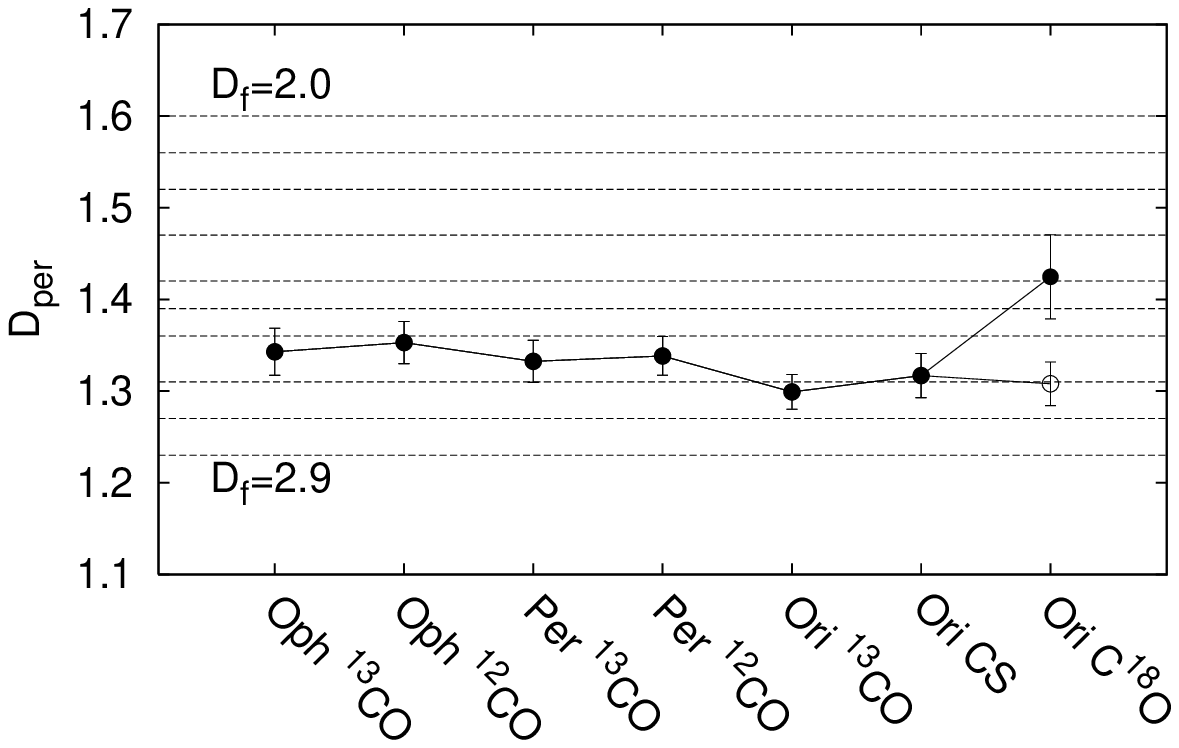}
\caption{The perimeter dimension $D_{per}$ obtained for each
molecular cloud map. The dashed horizontal lines indicate
values calculated in \citetalias{san05} for fractal dimension
values $D_f$ from 2.0 to 2.9 in increments of 0.1. The open
circle refers to the result obtained for the smoothed
C$^{18}$O map (see text).}
\label{perimeter}
\end{figure}

\begin{figure}
\plotone{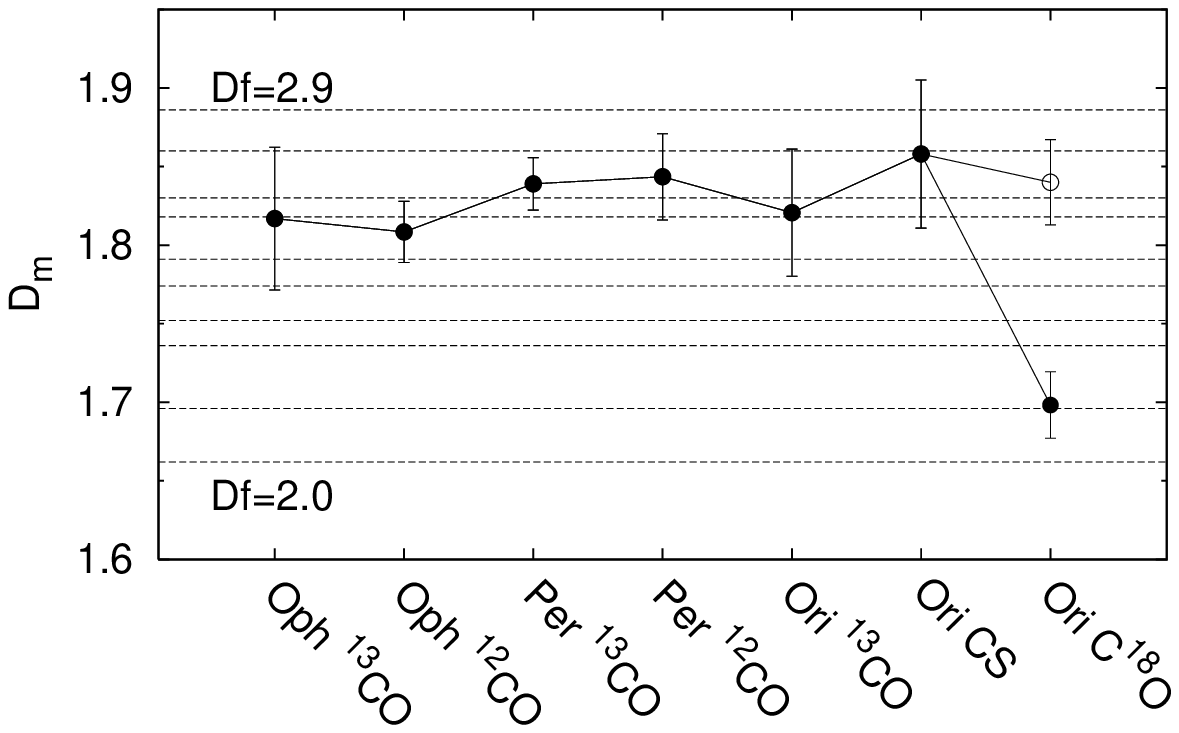}
\caption{The mass dimension $D_m$ obtained for each
molecular cloud map. The dashed horizontal lines indicate
values calculated in \citetalias{san05} for fractal
dimension values $D_f$ from 2.0 to 2.9 in increments
of 0.1. The open circle refers to the result obtained
for the smoothed C$^{18}$O map (see text).}
\label{mass}
\end{figure}

The results for the C$^{18}$O map of Orion are shown
in Figures~\ref{perimeter} and \ref{mass} as filled circles.
The $D_{per}$ value has a higher error bar for the C$^{18}$O
map of Orion, i.e., there is a worse correlation between
the perimeter and the area of the projected clouds. Moreover,
the resulting fractal dimension is in the range $D_f \simeq
2.3-2.5$, significantly lower than in the other maps. The
mass dimension also indicates a relatively low fractal
dimension but in this case the value is $D_f \simeq 2.0-2.2$.
In principle this would imply that the observed structures are
more irregular in the C$^{18}$O map than in the $^{13}$CO
y CS maps, but two points have to be taken into account
before coming to this conclusion.
First, the $D_f$ values derived from both estimators
($D_{per}$ and $D_m$) do not agree. Secondly, if the
C$^{18}$O map shows mainly dense regions where turbulence
is overcome by gravity in order to condense into prestellar
cores \citep{lar05} then the resulting structures should
be more regular, i.e., with higher fractal dimension values
\citep{fal04}.
Since the C$^{18}$O emission is much weaker than the other
ones the signal to noise ratio (S/N) is much lower for this
map. \citet{vog94} used Brownian fractals to show that noise
distorts the contours and thus tends to increase the estimate
of $D_{per}$. This is specially true in maps with low S/N
values. Thus, the results $D_{per} \sim 1.4$ and $D_m \sim 1.7$
for the C$^{18}$O map could be simply due to the fact
that very noisy maps produce more irregular structures, and
not necessarily meaning that C$^{18}$O is distributed in a
more irregular pattern in Orion A.
In other words, we have to try to disentangle structural
aspects from noise effects based only on the two-dimensional
projection of the cloud.

\begin{figure}
\plotone{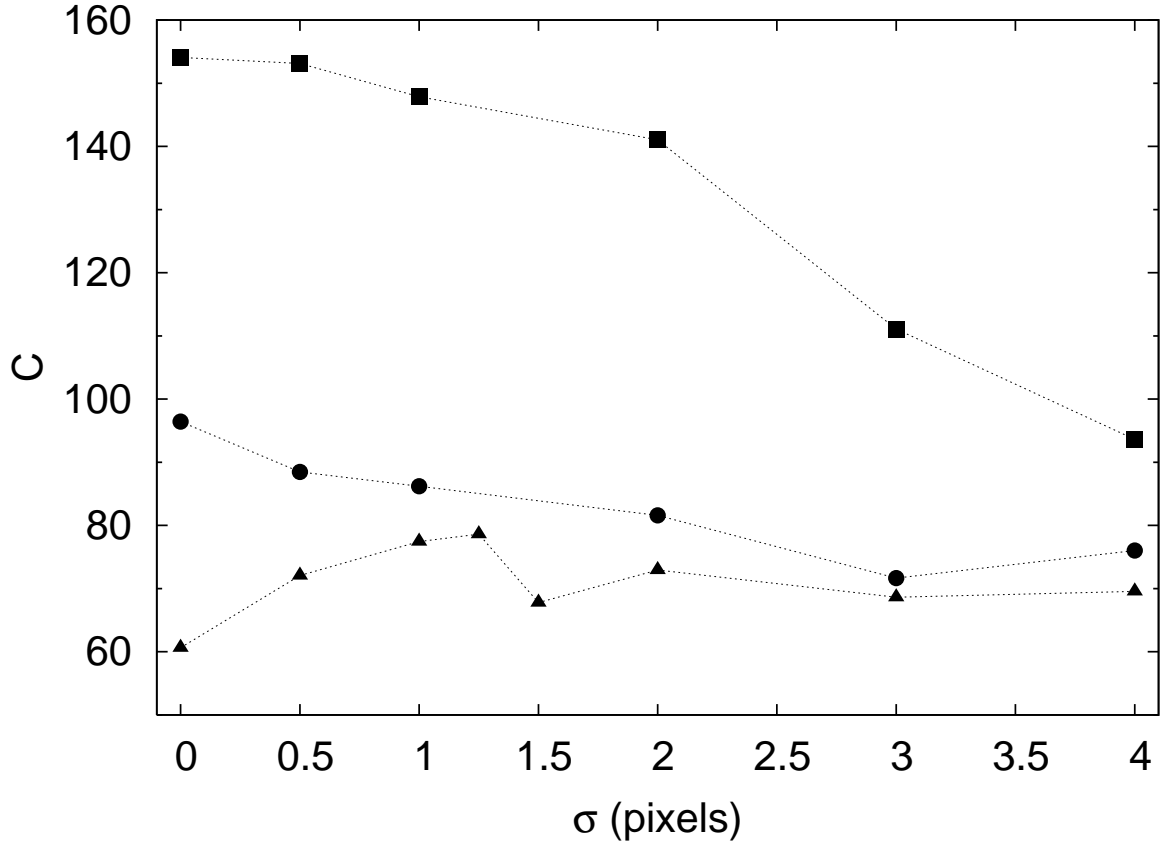}
\caption{The contrast parameter $C$ as a function of the
smoothing parameter $\sigma$ for the three maps of Orion
A used in this work: $^{13}$CO (squares), CS (circles) and
C$^{18}$O (triangles).}
\label{sigma}
\end{figure}

\begin{figure}
\plotone{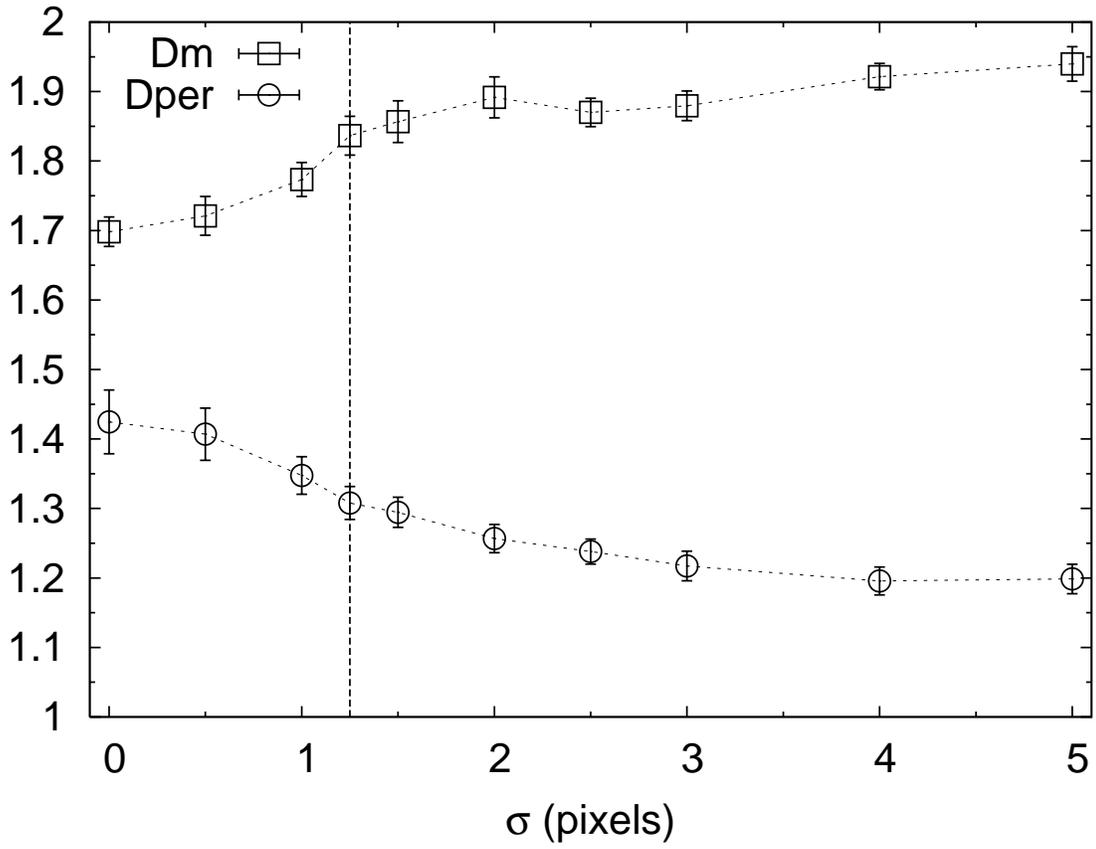}
\caption{The mass dimension (squares) and the perimeter
dimension (circles) as a function of the smoothing parameter
$\sigma$ for the C$^{18}$O map. The dashed vertical line
emphasizes the value at which the contrast ($C$) is maximum.}
\label{sigma2}
\end{figure}

In order to test this possibility, we proceeded to
increase the S/N ratio by smoothing the maps and then to
recalculate the fractal dimensions. We have used
a gaussian kernel to convolve the data, where the $\sigma$ of
the gaussian determines the size of the neighboring region
used to smooth spatial variations. If this variations between
neighbor pixels are due, in good part, to noise then the final
effect will be some reduction in the image noise level. An
optimal algorithm would maximize the S/N ratio throughout the
map, such as, for example, the adaptive kernel algorithms do
\citep{lor93,ebe06}.
Here we have used a simple space-invariant gaussian kernel
($\sigma$ constant) and we have calculated $D_{per}$ and
$D_m$ for different $\sigma$ values. In order to quantify
the contrasting quality of the resulting images after
smoothing we have introduced a new parameter $C$, named
``contrast", which takes into account the dynamical range of
the image and the rms of the background. This parameter is
defined as the ratio between the maximum intensity
in the map and the standard deviation of the
intensity values of the background
pixels. The calculation of $D_{per}$ is done by taking
a fixed number of brightness levels and finding all the
connected pixels (objects) whose brightness values
are above each predefined level \citepalias{san05}.
We consider here as ``background" pixels all pixels
whose brightness are below the minimum brightness level
considered to calculate $D_{per}$ (5\% of the maximum
brightness in the map). Thus, the parameter $C$ estimates
the contrast between the signal of the brightest object
in the map and the variations of the background pixels.
This parameter would be related to the S/N of the brightest
pixel only if the variations of the background pixels are
due mainly to noise. We have calculated $C$ for the original
maps and for the maps smoothed with different $\sigma$
values. Figure~\ref{sigma} shows the results for the three
maps of Orion A used in this work.\footnote{The Ophiuchus
and Perseus maps behaved similar to the $^{13}$CO and CS
maps of Orion A. For clarity those results are not presented
here.} As expected for a low S/N map, the C$^{18}$O map has
the lowest contrast, but the interesting result is that
this map is the only one that begins increasing $C$ as
$\sigma$ increases. This
means that as the map is smoothed the rms of the background
decreases faster than the peak intensity does.
In all the other maps the smoothing of
the background variations is accompanied by a decrease of the
maximum signal in a higher proportion. The contrast $C$ for
the C$^{18}$O map exhibits a maximum at $\sigma = 1.25$ pixels
(see Figure~\ref{sigma}).
This maximum represents the ``optimal" map, in the
sense of exhibiting the maximum contrast or, in other words,
the minimal noise distorion on the image (in the case that
background rms is due mainly to noise).
In each case we calculated $D_{per}$ and $D_m$ for the smoothed
maps and all the results showed the expected behavior, i.e.,
$D_{per}$ decreases (less irregular boundaries) and $D_m$
increases (more homogeneous distribution of intensities) as
$\sigma$ increases. Figure~\ref{sigma2} shows this result for
the C$^{18}$O map. For the $^{13}$CO and CS maps the same
behavior could be appreciated for $\sigma \gtrsim 1.5$
whereas for lower $\sigma$ values $D_{per}$ and $D_m$
remain more or less constant (within error bars).
The perimeter-area based dimension of the C$^{18}$O map for
the $\sigma$ at which $C$ reaches its maximum value
(shown as a vertical line in Figure~\ref{sigma2})
is $D_{per} = 1.31 \pm 0.02$ (shown as an open circle
in Figure~\ref{perimeter}), from which it is derived
that $D_f \simeq 2.7 \pm 0.1$, in very good agreement
with the results obtained from the $^{13}$CO and CS maps
(Figure~\ref{perimeter}). In addition, the mass dimension
for this case (maximum $C$ value)
yields $D_m = 1.84 \pm 0.03$ (open circle in
Figure~\ref{mass}) which again is consistent
within the error bars with the previous results.

To calculate $D_{per}$ we use a given number of intensity
levels in all the range of map intensities \citepalias{san05}.
In principle we expect that structure information in most of
these levels is only slightly distorted by noise in high S/N
maps. The lowest levels are probably more affected by noise,
but the perimeter dimension is calculated by using all the
objects in all the levels and therefore noise affects very
little the final result. The opposite occurs in low S/N
maps, where most of intensity levels are close to the noise
level and cloud boundaries may be artificially lengthened
(higher $D_{per}$ values).
The smoothing process should correct this problem
by flattening the wiggles due to noise in neighboring
pixels. But if the image is excessively smoothed the
clouds will exhibit unrealistic low $D_{per}$ values.
How much does the image have to be smoothed? In high S/N
images the distortion produced by noise is minimal, then
it is reasonable to impose the condition of maximizing
S/N in low S/N maps as a previous step in the estimation
of the fractal dimension. While this requirement
does not guarantee that the dimension obtained
is the ``real" one, it does ensure the ``best" estimation,
diminishing the effect of noise.
Since S/N may be an unknown quantity (besides
depending on the position in the map) we have looked
for a parameter connected to S/N but also easy to
calculate for a given map. The contrast $C$ defined
in this work equals the S/N of the
brightest pixel if the background
variations are due to noise. What we are suggesting
is that the results obtained when $C$ is maximum are
more ``realiable" than the results for the original
(unsmoothed) map.
These arguments are supported by the fact that both
estimators ($D_{per}$ and $D_m$) approach the same
$D_f$ value for the C$^{18}$O map when $C$ is a
maximum, and by the fact that this value agrees
with the other map values.

\section{CONCLUSIONS}
\label{sec_conclusions}

Both the perimeter dimension ($D_{per}$) and the mass
dimension ($D_m$) are useful tools to infer the
three-dimensional structure of molecular clouds from
two-dimensional maps. In general, $D_{per}$ yields
uncertainties smaller than $D_m$, but this last method
could be very useful to corroborate values and trends
observed in optically thin regions. The opacity does
not alter the results derived from the perimeter-area
method, but when $\tau \gtrsim 1$ the mass-size method
cannot be used in a reliable way to estimate the fractal
dimension $D_f$. An important point that should be
considered when using these methods with real data
is that very high noise levels can seriously affect
the estimation of $D_f$, decreasing artificially its
value. One possible strategy to prevent this situation
is the use of a smoothing algorithm that maximizes
the signal-to-noise ratio (S/N) throughout the map.
In this work we have defined a parameter called
``contrast" ($C$), which we propose can help to choose
the most ``reliable" image for estimating $D_f$.

From different emission maps of Ophiuchus, Perseus and
Orion molecular clouds we obtain that the fractal
dimension is always in the range $2.6 \lesssim D_f
\lesssim 2.8$. This result supports our previous
suggestion \citepalias{san05,san06} of a relatively
high ($> 2.3$) average fractal dimension for the ISM.
The ultimate goal is to understand the origin of the
ISM structure, therefore it would be important to investigate
what physical processes are able to generate high fractal
dimension structures.

\acknowledgments
We want to thank the referee for his/her helpful
comments and criticisms, which improved this paper.
N.~S. would like to acknowledge funding provided by the
Secretar\'{\i}a de Estado de Universidades e Investigaci\'on
(Spain) through grant SB-2003-0239.
E.~J.~A. acknowledges funding from MECyD of Spain through
grants AYA2004-05395 and AYA2004-08260-C03-02, and from
Consejer\'{\i}a de Educaci\'on y Ciencia (Junta de Andaluc\'{\i}a)
through TIC-101.
E.~P. acknowledges financial support from grants
AYA2004-02703 and TIC-114.

\clearpage
\begin{deluxetable}{ccccc}
\tablecolumns{5}
\tablewidth{0pt}
\tablecaption{Calculated fractal dimension\label{table1}}
\startdata
\cutinhead{Perimeter-area based dimension ($D_{per}$)}
$D_f$ & $\tau_0=0.0$ & $\tau_0=1.0$ & $\tau_0=2.0$ & $\tau_0=5.0$ \\
\tableline
2.0 & 1.601$\pm$0.024 & 1.602$\pm$0.021 & 1.604$\pm$0.021 & 1.591$\pm$0.019 \\
2.3 & 1.469$\pm$0.023 & 1.474$\pm$0.021 & 1.467$\pm$0.019 & 1.455$\pm$0.018 \\
2.6 & 1.359$\pm$0.032 & 1.364$\pm$0.032 & 1.367$\pm$0.035 & 1.359$\pm$0.046 \\
\cutinhead{Mass dimension ($D_m$)}
$D_f$ & $\tau_0=0.0$ & $\tau_0=0.5$ & $\tau_0=1.0$ & $\tau_0=1.25$ \\
\tableline
2.6 & 1.808$\pm$0.029 & 1.816$\pm$0.034 & 1.876$\pm$0.045 & 1.860$\pm$0.044 \\
\enddata
\end{deluxetable}

\clearpage
\clearpage
\clearpage
\clearpage
\clearpage
\clearpage

\end{document}